# Investigation of temperature stress tolerance in Arabidopsis STTM165/166 using electrophysiology and RNA-Seq

Dongjie Zhao[1*], Qinghui Chen[2], Ziyang Wang[3], Lucy Arbanas[2], Guiliang Tang[2*]

[1]Institute for Future (IFF), School of Automation, Qingdao University, Qingdao, 266071, China
[2]Department of Biological Sciences, Michigan Technological University, Houghton, Michigan, 49931, USA
[3]Institute of Automation, Chinese Academy of Sciences, Beijing, 100190, China

***Correspondence:**
Dongjie Zhao

dongjiezhao@qdu.edu.cn

Guiliang Tang

gtang1@mtu.edu

## Abstract

Plant electrical signals have been shown to be generated in response to various environmental stresses, but the relationship between these signals and stress tolerance is not well understood. In this study, we used the Arabidopsis STTM165/166 mutant, which exhibits enhanced temperature tolerance, to examine this relationship. Surface recording techniques were utilized to compare the generation ratio and duration characteristics of electrical signals in the STTM165/166 mutant and wild type (WT). Patch-clamp recording was employed to assess ion channel currents, specifically those of calcium ions. The current intensity of the mutant was found to be lower than that of the WT. As calcium ions are involved in the generation of plant electrical signals, we hypothesized that the reduced calcium channel activity in the mutant increased its electrical signal threshold. RNA-Seq analysis revealed differential expression of AHA genes in the STTM165/166 mutant, which may contribute to the prolonged depolarization phenotype. Gene Ontology enrichment of differentially expressed genes (DEGs) identified associations between these DEGs and various stresses, including temperature, salt, and those related to the jasmonic acid and abscisic acid pathways. These findings provide experimental evidence for the use of plant electrical signals in characterizing stress tolerance and explore potential ion mechanisms through patch-clamp recording and DEG Gene Ontology analysis. They also emphasize the need for further research on the relationship between plant electrical signals and stress tolerance.

## Introduction

Environmental stresses are an integral part of plant growth and development. Stress tolerance is vital for adaptive growth and increased survival under adverse conditions (Vishwakarma et al., 2017). Plant electrical signals are the plant's initial response to environmental stress, involving the rapid transmission of information through changes in cell membrane potential caused by transmembrane ion movement (Fromm and Lautner, 2007; Hedrich, 2012). Many studies have shown that plant electrical signals can be induced by various environmental stresses, such as temperature, drought, and salt stress. For example, Sukhov et al. applied high-temperature stress to peas and recorded plant electrical signals in the leaves, revealing that the frequency of electrical activity was related to the intensity of stress (Sukhov et al., 2017). Fromm et al. recorded low-temperature stress-induced electrical signals, primarily action potentials, in maize leaves (Fromm et al., 2013). Vuralhan-Eckert et al. recorded electrical activity in response to drought stress applied to plant roots and found it could regulate the CO2 exchange rate and photosynthetic efficiency in leaves (Vuralhan-Eckert et al., 2018; da Silva et al., 2020). Stolarz et al. recorded salt stress-induced long-distance electrical activity in plant stems and leaves, showing that stress can regulate spontaneous action potentials (Stolarz and Dziubinska, 2017). These findings underscore the importance of further investigating the relationship between plant electrical signals and stress tolerance.

Signal processing techniques are frequently used to analyze plant electrical signals. The analysis of these signals typically focuses on the dynamic mechanisms of signal generation and transmission, feature extraction, and classification. For instance, Sukhov et al. developed a mathematical model of the generation and conduction of plant electrical signals under environmental stress (Sukhov and Vodeneev, 2009; Ekaterina et al., 2017). Wang et al. used the modified Hodgkin-Huxley equation to describe the process of cell membrane potential change based on published experimental data (Wang et al., 2007). Huang et al. utilized independent component analysis to analyze action potential and variation potential, discovering the independent components of surface recording electrical signals (Huang et al., 2009). Chatterjee et al. classified plant electrical signals according to stress stimulation type using nonlinear, naive Bayesian linear, and quadratic discriminant analysis (Chatterjee et al., 2015). Qin et al. found that a one-dimensional convolutional neural network could extract the features of wheat electrical signals under stress (Qin et al., 2020). These studies demonstrate the usefulness of signal processing methods in understanding plant electrical signals and their relationship to stress.

RNA sequencing (RNA-seq) is a powerful tool for identifying differentially expressed genes (DEGs) and understanding gene regulatory networks in response to stress. For example, Ghaffari et al. used RNA-seq to identify DEGs in Arabidopsis thaliana in response to salt stress and revealed a significant role for the transcription factor WRKY70 in the stress response (Ghaffari et al., 2015). Similarly, Liu et al. used RNA-seq to identify DEGs in maize in response to drought stress and found that the transcription factor ZmMKK4 played a key role in the stress response (Liu et al., 2016).

Electrical signals can serve as important indicators of an organism's physiological state through the extraction of their characteristics. For instance, animal EEG and ECG signals have been widely used to reflect brain activity and heart disease. Similarly, plant electrical signals may be a means of reflecting a plant's physiological state, particularly its stress tolerance. However, few studies have focused on using plant electrical signals to characterize stress tolerance in plants. In this study, we used Arabidopsis STTM165/166 mutants, constructed using a short tandem target mimic (Yan et al., 2012), which exhibit higher temperature stress tolerance than the wild type, to investigate the relationship between electrical signal phenotype and stress resistance and to explore the ion mechanisms underlying this relationship. We also identified potential gene regulatory networks that play a crucial role in this process using RNA-seq to identify DEGs in the mutants. Our findings demonstrate the potential of using plant electrical signals as indicators of temperature stress tolerance and highlight the importance of further investigating the relationship between plant electrical signals and environmental stresses.

## Materials and Methods

### Plant Materials and Culture Conditions

Arabidopsis thaliana plants of the Columbia-0 ecotype were used in this study, including wild type (WT) and STTM165/166 mutants. The STTM165/166 mutant was created using short tandem target mimic technology (Yan et al., 2012). Seeds were sterilized using 70% ethanol for 1 min, followed by 5% bleach for 5 min, and then thoroughly rinsed with sterilized water. The sterilized seeds were then sowed on 1/2 Murashige and Skoog medium (1/2MS) with a pH of 5.8 and 0.8% agar, and incubated at 4°C for 3 days to promote germination. The seedlings were then transferred to sterilized soil under a 14 h light/10 h dark photoperiod and watered every 2 days. The humidity was maintained at 70%, and the temperatures were 22°C (day) and 18°C (night).

### Plant Electrical Signal Surface Recording

For surface potential recording, 3- to 4-week-old plants were used. The experiments were conducted in a controlled environment room, under the same growth conditions as before. Ag/AgCl electrodes (0.5 mm diameter silver wire, chloridized with HCl 0.1 M) were used for recording (World Precision Instruments, Sarasota, FL). The electrode-leaf interface was established using a drop (10 µL) of 10 mM KCl in 0.5% (w/v) agar. The ground electrode was placed in the soil. A heat or cold stimulus (approximately 0.5 cm3 of ice or 0.5 cm3 of 45°C heat water, sealed in a plastic bag) was applied to the leaf tip, with the recording electrode placed about 1.5 cm from the stimulus region (Figure 1a). The stimulus was maintained on the leaf during the recording time (approximately 300 s). The output voltage of the surface recording was passed through a high-impedance operational amplifier (FD 223, World Precision Instruments, Sarasota, FL) used as a voltage follower. The signal was digitized and stored in a signal acquisition system (LabTrax 4-Channel Data Acquisition, World Precision Instruments, Sarasota, FL) using Datatrax2 software. The plant was placed on an anti-vibration table in a Faraday cage (Ametek TMC, Peabody, MA). To statistically compare WT and STTM165/166, four parameters were used to characterize the electrical signal: depolarization time (T1), repolarization time (T2), duration, and integral area (Figure 1b). The statistical values of T1, T2, duration, and integral area are presented as mean ± SD.

### Plant Protoplast Isolation

3- to 4-week-old plant protoplasts were used for patch clamp recording. The protoplast isolation method has been described previously (Lemtiri-Chlieh and Berkowitz, 2004). Briefly, epidermal strips of leaves containing guard cells were floated on a medium containing 1.8-2.5% (w/v) Cellulase Onozuka RS (Yacult Honsha, Tokyo, Japan), 1.7-2% (w/v) Cellulysin (Calbiochem, Behring Diagnostics, La Jolla, CA), 0.026% (w/v) Pectolyase Y-23 (Yacult Honsha), 0.26% (w/v) bovine serum albumin, and 1 mM CaCl2 (pH 5.6) with osmolality adjusted to 360 mOsm/kg using mannitol. The protoplasts were released after 2-3 h of incubation at 28°C with gentle shaking, passed through a 25-µm mesh, and kept on ice for 2-3 min before centrifugation (100x g for 4 min at room temperature). For mesophyll cell protoplasts, the peeled leaves and tissue were incubated for 20-30 min before centrifugation. The pellet of guard cell or mesophyll cell protoplasts was resuspended and kept on ice in 1-2 mL of fresh medium containing 0.42 M mannitol, 10 mM Mes, 200 µM CaCl2, 2.5 mM KOH (pH 5.55 and osmolality at 466 mOsm/kg). Unless stated otherwise, all chemicals were from Sigma.

## Protoplast Patch Clamp Recording

In this research, patch clamp recording was performed using a protocol previously described (Lemtiri-Chlieh and Berkowitz 2004). The focus of the recording was on calcium current. To record the calcium current, barium-containing solutions were used, containing 50 mM BaCl2, 1 mM KCl, and 10 mM Mes (pH 5.5 with KOH) for the bath, and 5 mM BaCl2, 20 mM KCl, and 10 mM Hepes (pH 7.5 with KOH) for the pipette. The osmolality of the solutions was adjusted to 470 mOsm/kg (bath) or 500 mOsm/kg (pipette) with mannitol. The cAMP was solubilized in deionized water and stored in aliquots of 50-100 μl at a concentration of 0.1 M. A few minutes before the experiment, the cAMP solution was diluted to the final desired concentration (1 mM). After preparing the solutions, 20 mL of protoplasts were placed in a 2-mL chamber and continuously perfused at a flow rate of 0.5 mL/min.

Patch pipettes with diameters of 5-10 μm were pulled from Kimax-51 glass capillaries (Kimble 34500; Kimble, Owens, IL) using a Flaming/Brown micropipette puller (Sutter P-97; Sutter Instrument Company, Novato, CA). The experiments were conducted at room temperature (20-22 °C) using standard whole-cell patch clamp techniques, with an Axopatch 200B integrating patch clamp amplifier (Axon Instruments, Inc., Union City, CA). Voltage commands and simultaneous signal recordings and analyses were performed using a microcomputer connected to the amplifier via a Digidata 1320A multipurpose input/output device and pClamp 9.0 software (Axon Instruments, Inc.). After forming gigaohm seals in the cell-attached configuration, the whole-cell configuration was achieved by gently applying suction to the membrane, which was then clamped to a holding potential of -30 mV. The standard voltage protocol ramped from 50 mV to -200 mV over a period of 2 seconds. In all experiments, there was a waiting time of 3-5 minutes following the establishment of the whole-cell configuration before any current measurements were made. All current traces were low-pass filtered at 2 kHz before analog-to-digital conversion.

## RNA Sequencing and Analysis of the Data

RNA sequencing was performed using a previously described method (Gu, 2017). Total RNA was extracted from 3-week-old wild-type and STTM165/166 transgenic plant leaves using Trizol reagent (Invitrogen, MA, USA) and the quality was assessed by agarose gel electrophoresis and RNA Integrity Number (RIN) analysis with a Bio-analyzer. cDNA libraries were constructed and sequenced on the Illumina HiSeq platform at the University of Michigan DNA sequencing core facility. Differential expression analysis was performed using the DESeq method, where genes with an adjusted P-value $< 0.05$ and a fold change greater than or less than 1/threshold (where threshold = 1.5) were considered differentially expressed. The differentially expressed genes were then analyzed for their involvement in biological processes using Gene Ontology (http://www.geneontology.org/).

# Results

## Electrical Signal Response to Temperature Stress in Arabidopsis STTM165/166

The typical recording of the cold shock-induced electrical signal is shown in Figure 1c. Both single action potentials (APs) and series of APs were recorded in wild-type and STTM165/166 plants. Under heat shock, we only recorded a single AP, the typical signal of which is shown in Figure 1d. In comparing the parameters of the cold shock-induced electrical signals, we found little difference in the integral area of single APs or series of APs and T1 time (Figure 1e and 1f). However, the mutant showed prolonged duration and T2 time compared to the wild-type (Figure 1f). Additionally, the generation ratio of electrical activities was higher in the wild-type than in the mutant (Figure 1g). The integral area and T1 time comparison for heat shock-induced APs showed similar results to those for cold shock-induced APs, with little difference (Figure 1h). In addition, we found that the duration time and repolarization time T2 were longer in the mutant than in the wild-type (Figure 1i).

**Calcium Channel Activity in Arabidopsis STTM165/166 Under Temperature Stress**

The surface potential phenotype is influenced by the transmembrane movement of calcium, potassium, and chloride ions. From RNA-seq analysis, we identified differentially expressed genes (DEGs) related to these ions. Most ion channel genes did not show significant expression differences (Supplementary Material, Tables S1-S5), with the majority of DEGs belonging to the cyclic nucleotide-gated channel (CNGC) family (Table 1). CNGCs, a 20-member family in Arabidopsis, are thought to mediate $Ca^{2+}$ signals in plants.

To verify whether there is a difference in calcium channel activity, patch clamp techniques were used to record calcium currents in wild-type and mutant plants. Figure 2a shows the recording results for wild-type Arabidopsis thaliana guard cells. When the channel activator cAMP was absent, a small current of approximately -20 pA was recorded at the -200 mV holding potential. The ionic current intensity increased significantly when cAMP was added to the cell solution, reaching approximately -100 pA at -200 mV. When the calcium channel blocker $LaCl_3$ was added to the solution, the current intensity decreased significantly, indicating that the activated current was primarily a calcium current. Figure 2b shows the recording results for mutant guard cells, which displayed similar trends to those in Figure 2a. When cAMP was absent, the recorded current was approximately -20 pA. When cAMP was added, the current increased to approximately -80 pA at -200 mV clamping voltage. Figure 2c shows the recording of wild-type Arabidopsis thaliana mesophyll cells, which displayed similar trends to those in guard cells. When cAMP was absent, the current was approximately -30 pA. The current increased significantly to approximately -250 pA when cAMP was added. The current in mutant mesophyll cells was much smaller, only reaching approximately -100 pA at the -200 mV holding potential.

**Differentially Expressed Genes and Their Physiological Processes in Arabidopsis STTM165/166 Under Temperature Stress**

From STTM165/166 RNA-seq analysis, we found DEGs in the CNGC gene family and verified the DEGs' influence on the calcium current by patch clamp recording. Besides the DEGs in the CNGC family, the RNA-seq analysis revealed more than 3000 DEGs. For these DEGs, the number of upregulated genes was significantly more prominent than that of down-regulated. To further explore the relationship between these DEGs and environmental stress. We figured out the DEGs biological process by GO tool, as shown in figure 3. Figure3(a) is the biological process map of DEGs.We found that nearly half of the biological processes are associated with environmental stress factors, such as cold, heat, salt etc. And the processes that respond to the plant hormones, jasmonic acid, abscisic acid, and salicylic acid, also have a big proportion. Figure3(b) lists the response or defense-related biological processes separately by upregulated and downregulated genes. We found that the upregulated genes regulate much more numbers of biological processes than downregulated genes. And the biological processes as response to cold, response to heat, response to salt stress, response to abscisic acid, response to jasmonic acid etc., each contains more than 50 upregulated genes. Furthermore, these biological processes are not isolated, and there are complex links between them, as shown in figure3(c).

### Exploration of the Regulatory Network among DEGs in Arabidopsis STTM165/166 under Stress Conditions

We identified a selection of DEGs that are involved in regulated physiological processes and their interconnections (Figure 3). Using the regulatory network among these DEGs as a foundation, we were able to understand how these processes are connected. Specifically, we found that the regulatory network serves as the foundation for this network of physiological processes, allowing us to better understand the connections between these DEGs as the following.

Upon such analysis, we observed a portion of the DEGs regulatory network starting with the target genes of mir165/166 (Figure 4). These target genes, including PHABULOSA (PHB), PHAVOLUTA (PHV), REVOLUTA (REV), ATHB-8, and ATHB-15, were found to be connected to other regulatory networks through the use of bridging genes, such as Ago10, MYBb51, IMK2, and others. These bridging genes acted as connectors, linking various regulatory networks to the second shell. For example, the cyclic nucleotide-gated channel (CNGC) network, the response to jasmonic acid (JA) network, the response to abscisic acid (ABA) network, and the response to salt stress network were all linked to the second shell through WRKY33, MYB51, S6K2, IMK2, MYC2, and other genes. In addition, we observed connections to other networks, such as the response to cold and salicylic acid, through these bridging genes. These findings provide insight into the complex regulatory network among DEGs and the physiological processes it controls.

## Discussion

### Electrical Signal and Calcium Channel Dynamics in Arabidopsis STTM165/166 Under Temperature Stress

In this study, we examined the electrical signal and calcium channel dynamics in Arabidopsis STTM165/166 plants under temperature stress. Our results showed that the duration and repolarization time of cold shock-induced electrical signals were prolonged in the mutant plants compared to the wild-type. Similarly, heat shock-induced electrical signals displayed longer durations and repolarization times in the mutants. These findings suggest that the mutant plants may have altered electrical signal responses to temperature stress compared to the wild-type, potentially impacting their ability to withstand or recover from such stressors.

In addition to examining electrical signals, we also investigated calcium channel activity in the mutant and wild-type plants. Patch clamp recordings revealed similar trends in both guard cells and mesophyll cells, with the current intensity increasing in the presence of the channel activator cAMP (Li et al., 2019). However, the mutant mesophyll cells displayed significantly smaller current intensities compared to the wild-type (Li et al., 2019). These results indicate that there may be differences in calcium channel activity between the mutant and wild-type plants under temperature stress, potentially impacting the plants' calcium signaling pathways and related responses to stress (Sukhova et al., 2020).

Overall, our results highlight the potential role of electrical signals and calcium channel dynamics in the response of Arabidopsis STTM165/166 plants to temperature stress (Fromm and Lautner, 2007). Understanding these mechanisms is important for improving our understanding of plant stress responses and developing strategies for improving plant stress tolerance (Krausko et al., 2017; Hedrich and Neher, 2018; Li et al., 2019). Additionally, analyzing the differentially expressed genes identified in this study may provide further insight into the molecular mechanisms underlying these stress responses and how they may be manipulated to enhance plant stress tolerance.

## Variation in Plant Electrical Signals

There is a wide range of variation in the structure and duration of plant electrical signals, as observed in numerous experimental studies (Fromm and Lautner, 2007). These variations can be influenced by various factors such as plant species, environmental stresses, and plant stress resistance. For example, the Venus flytrap and sundews display rapid depolarization and repolarization processes that last only a few seconds (Krausko et al., 2017; Hedrich and Neher, 2018). In other plant species such as broad bean, cucumber, Zea mays, Nicotiana tabacum, summer squash, sunflower, and barley, recorded electrical signals have durations ranging from a few seconds to several minutes (Fromm and Fei, 1998; Felle and Zimmermann, 2007; Zimmermann and Felle, 2009; Sukhova et al., 2020). There are also differences in electrical signals for the same plant species under different types of stress, including biotic and abiotic stresses, non-wounding stimuli, and wounding stimuli. For example, in Arabidopsis thaliana, electrical signals generated in response to non-wounding stimuli tend to have shorter durations compared to those in response to wounding stimuli (Hilleary and Gilroy, 2018; Kollist et al., 2019; Farmer et al., 2020). When the stimulus is kept constant for a particular plant species, the differences in electrical signal phenotype may be primarily due to variations in plant stress resistance. For example, when exposed to controlled light-dark stimuli, wheat plants with different levels of stress resistance displayed varying amplitudes and durations in their electrical signals (Li et al., 2019). We also observed distinct electrical signal phenotypes in STTM165/166 Arabidopsis mutants compared to wild-type plants.

## Role of Electrical Signals in Stress Resistance

In many studies, the STTM165/166 Arabidopsis mutant has been proven with higher stress resistance than the wild type. For example, the STTM165/166 was shown to have a higher survival rate under salt stress or drought stress (Peng et al., 2018; Yang et al., 2019). In this study, we found that the mutant also had higher resistance under temperature stress (Supplementary Figure S1). Based on the different stress resistance of the mutant and the wild type, the recorded electrical signal phenotypes showed their differences (Figure 1). The mutant generally has a smaller generation probability than the wild-type. In addition, the mutant has a longer duration time, especially repolarization time, than the wild-type under cold and heat stress. These results indicate that for the higher resistant plant STTM165/166, the stress-induced electrical signal has a slower response, with a longer duration and lower generation ratio. These findings suggest that the sensitivity of plant electrical signals may be inversely proportional to the stress resistance of plants. In other words, plants with higher environmental stress resistance may have stress-induced electrical signals with higher impedance to generation and waveform evolution.

## Ion Channels in Plant Electrical Signals

Many studies have investigated the ion mechanism of plant electrical signals, and different ion channels have been identified in plants (Hedrich, 2012). In general, for plant action potentials usually induced by non-noxious stress, their depolarization and repolarization are determined by the transmembrane movement of calcium, chloride, potassium ions, and protons. Current studies have shown that the initial depolarization of action potentials in plants is caused by calcium influx due to the opening of $Ca^{2+}$ channels, which leads to the opening of calcium-dependent $Cl^{-}$ channels, and the outflow of $Cl^{-}$ leads to further depolarization of the membrane potential. When the depolarization reaches a certain extent, voltage-dependent potassium channels will open, and then $K^{+}$ influx dominates the repolarization of the membrane potential. In addition to $Ca^{2+}$, $Cl^{-}$, $K^{+}$, proton pumps, and many transporters are involved in electrical activity. Compared to ion channels in animals, the response of ion channels in plants to environmental stress is relatively poorly understood. Our study found that under temperature stress, the mutant STTM165/166 had a higher expression of ion channels in the CNGC family compared to the wild-type. We also observed differences in calcium channel activity between the mutant and the wild-type using patch clamp recording techniques. These findings suggest that ion channels, particularly those in the CNGC family, may play a role in plant stress responses and electrical signaling.

In summary, our study found that the STTM165/166 Arabidopsis mutant had distinct electrical signal phenotypes and ion channel expression patterns under temperature stress compared to the wild-type. These findings suggest that plant electrical signals may be influenced by plant stress resistance and that ion channels, particularly those in the CNGC family, may play a role in plant stress responses and electrical signaling. Further research is needed to fully understand the mechanisms underlying these observations and the role of plant electrical signals in stress responses.

## The gene regulating network in Arabidopsis STTM165/166 under stress

In this study, we investigated the differentially expressed genes (DEGs) and their associated physiological processes in Arabidopsis STTM165/166. The regulatory network among the DEGs provided insights into the interconnectedness of physiological processes. This comprehensive analysis of the DEGs and their regulatory network provides clues into the complex interplay between genes and physiological processes in Arabidopsis STTM165/166 under stress conditions. It highlights the role of specific genes and their connections in orchestrating various stress responses, hormone signaling pathways, and other regulatory networks. Understanding these intricate relationships will contribute to a deeper understanding of plant stress adaptation and potentially facilitate the development of strategies for improving stress tolerance in crops. Further studies can focus on the functional characterization of specific genes and their roles in modulating interconnected physiological processes (Li et al., 2023). Overall, our findings shed light on the regulatory mechanisms underlying the response to temperature stress in Arabidopsis and provide a foundation for future research in unraveling the intricate network of genes and physiological processes in plants.

## Conclusions

In this study, we recorded temperature stress-induced electrical signals in Arabidopsis STTM165/166 mutants and wild-type plants. Our results showed differences in the generation and duration of the electrical signals between the mutant and wild-type plants. This suggests that the stress-induced electrical signals of plants with higher environmental stress resistance may have a higher impedance to generation and waveform evolution. To understand the ionic mechanisms underlying these differences in electrical signal phenotype, we conducted RNA-seq analysis and patch clamp recording. Our RNA-seq analysis revealed several biological processes, such as response to heat and cold, enhanced jasmonic acid and abscisic acid signaling, that contribute to the mutant's enhanced stress tolerance. Gene regulatory network analysis also identified several genes that link the electrical signal pathway with stress-related biological processes and serve as nodes in the electrical signal and stress resistance network.

Overall, our findings provide new insights into the complex regulatory network of differentially expressed genes and their role in physiological processes under temperature stress in Arabidopsis. Further studies are needed to fully understand the mechanisms behind these observations and the potential role of plant electrical signals in stress responses.

It is worth noting that temperature stress is just one of many types of environmental stresses that plants encounter. Future studies should explore the role of electrical signals in the response to other types of stresses, such as drought or herbivory. Additionally, it would be interesting to examine the relationship between stress resistance and electrical signaling in other plant species to see if the findings from this study are applicable more broadly.

In conclusion, our study adds to the growing body of evidence highlighting the importance of electrical signaling in plant stress responses. Understanding the mechanisms behind these signals and their role in stress resistance could have important implications for plant breeding and crop improvement efforts aimed at increasing plant stress tolerance.

## Conflict of Interest

The authors declare that there are no conflicts of interest regarding the publication of this manuscript. The research was conducted without any commercial or financial relationships that could be perceived as a potential conflict of interest.

## Author Contributions

D.J.Z., G.L.T., and Q.H.C. designed the experiments and D.J.Z. performed the experiments. D.J.Z., G.L.T., and Z.Y.W. analyzed the data, and D.J.Z. and G.L.T. wrote the manuscript. All authors participated in the discussions of the results and the preparation of the manuscript.

**Funding**

This research was funded by the Natural Science Foundation of Shandong province, China (grant number ZR201911050109) and the NSF (grant number 1340001).

**Acknowledgments**

We would like to express our gratitude to Professors Lan Huang, Zhongyi Wang and Shuzhi Sam Ge for their guidance and support throughout this research project. We also extend our thanks to Drs. Sachin Teotia, Lina Shi, and Haiping Liu for their assistance with plant material preparation. Additionally, we appreciate the help of Tao Liu with patch clamp recording.

Table 1 RNA-seq data of CNGC gene family

| test_id | gene | Descriptic | samp | sampl | status | value_ | value_2 | log2(fo | test_sta | p_value | q_value | Fold Chang | Diff Exp |
|---|---|---|---|---|---|---|---|---|---|---|---|---|---|
| AT5G15 | CNGC2 | cyclic nuc | G166 | WT | OK | 35.2 | 73.6 | 1.063 | 4.018 | 5E-05 | 6E-04 | 2.08866 | YES |
| AT2G24 | CNGC1 | cyclic nuc | G166 | WT | OK | 0.63 | 1.229 | 0.976 | 1.838 | 0.007 | 0.035 | 1.96682 | YES |
| AT2G46 | CNGC1 | cyclic nuc | G166 | WT | OK | 4.39 | 1.993 | -1.14 | -3.044 | 5E-05 | 6E-04 | 2.20447 | YES |
| AT2G46 | CNGC3 | cyclic nuc | G166 | WT | OK | 4.58 | 1.748 | -1.39 | -3.491 | 5E-05 | 6E-04 | 2.62168 | YES |
| AT3G17 | CNGC1 | cyclic nuc | G166 | WT | OK | 0.83 | 0.214 | -1.95 | -2.927 | 5E-05 | 6E-04 | 3.87297 | YES |
| AT1G19 | CNGC8 | cyclic nuc | G166 | WT | OK | 0.27 | 0.029 | -3.24 | -2.709 | 0.008 | 0.039 | 9.4457 | YES |
| AT5G57 | CNGC5 | cyclic nuc | G166 | WT | OK | 31.2 | 45.68 | 0.551 | 2.03 | 3E-04 | 0.002 | 1.4648 | NO |
| AT1G15 | CNGC7 | cyclic nuc | G166 | WT | OK | 0.13 | 0.263 | 1.066 | 0.126 | 0.777 | 0.888 | 2.093 | NO |
| AT5G14 | CNGC1 | cyclic nuc | G166 | WT | OK | 0.38 | 0.689 | 0.861 | 1.345 | 0.038 | 0.127 | 1.81668 | NO |
| AT2G46 | CNGC1 | cyclic nuc | G166 | WT | OK | 27 | 33.61 | 0.316 | 1.177 | 0.042 | 0.135 | 1.24486 | NO |
| AT1G01 | CNGC1 | cyclic nuc | G166 | WT | OK | 2.86 | 3.385 | 0.243 | 0.634 | 0.283 | 0.5 | 1.18312 | NO |
| AT5G53 | CNGC1 | cyclic nuc | G166 | WT | OK | 24.3 | 24.73 | 0.026 | 0.074 | 0.899 | 0.954 | 1.01826 | NO |
| AT5G54 | CNGC4 | cyclic nuc | G166 | WT | OK | 33.6 | 34.16 | 0.022 | 0.083 | 0.882 | 0.944 | 1.0156 | NO |
| AT2G23 | CNGC6 | cyclic nuc | G166 | WT | OK | 24.1 | 23.58 | -0.03 | -0.124 | 0.846 | 0.926 | 0.97982 | NO |
| AT2G28 | CNGC1 | cyclic nuc | G166 | WT | OK | 4.18 | 3.974 | -0.07 | -0.223 | 0.732 | 0.863 | 0.95008 | NO |
| AT4G30 | CNGC9 | cyclic nuc | G166 | WT | OK | 3.4 | 3.162 | -0.11 | -0.145 | 0.804 | 0.903 | 0.92889 | NO |

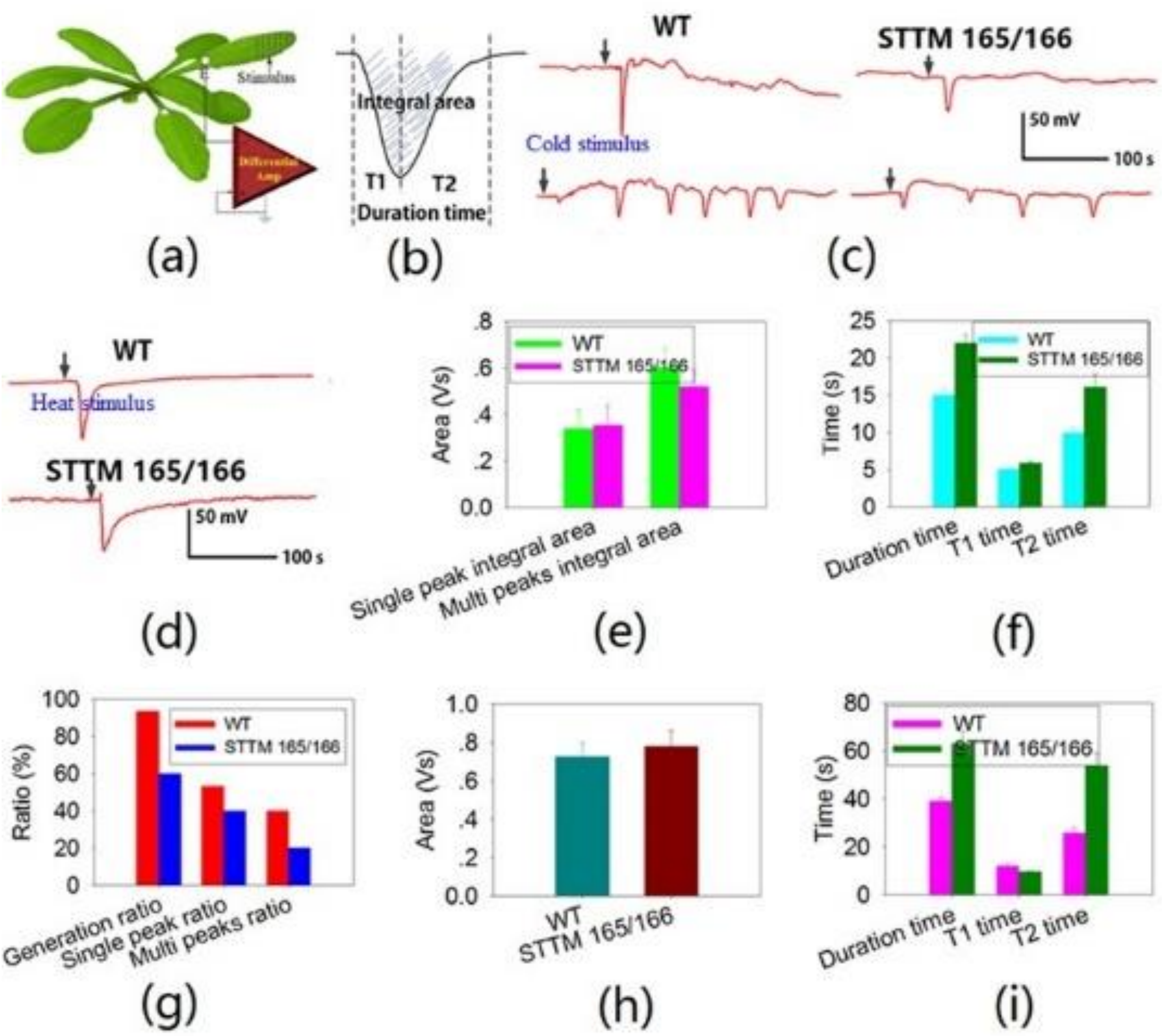

Figure 1. Suppression of Action Potential Induced by Temperature Stress in STTM 165/166. (a) In the action potential recording schematic, a stimulus (heat shock or cold shock) is applied to the tip of one leaf, and a recording electrode is placed on the same leaf. (b) The statistics show that the action potential (AP) induced by temperature stress is suppressed in STTM 165/166 when compared to the wild-type (WT) in terms of T1, T2, duration time, and the integral area. (c) Single AP or series of APs are induced by cold shock in both the WT and STTM 165/166. (d) Typical heat shock-induced action potential in wide type and STTM 165/166. (e) Integral area analysis of cold shock-induced AP or APs. (f) Duration time analysis of cold shock-induced AP and APs. In (e) and (f), WT n=14, STTM165/166 n=9. (g) Generation ratio of AP induced by cold shock. Wide type and STTM 165/166, n=15. (h) Integral area of heat shock-induced AP in wide type and STTM 165/166. (i) Duration time analysis of heat shock-induced AP. In (h) and (i), WT n=9, STTM 165/166 n=8. All statistical values are presented as the mean ± standard error of the mean (SE) and the sample sizes (n) are indicated for each comparison.

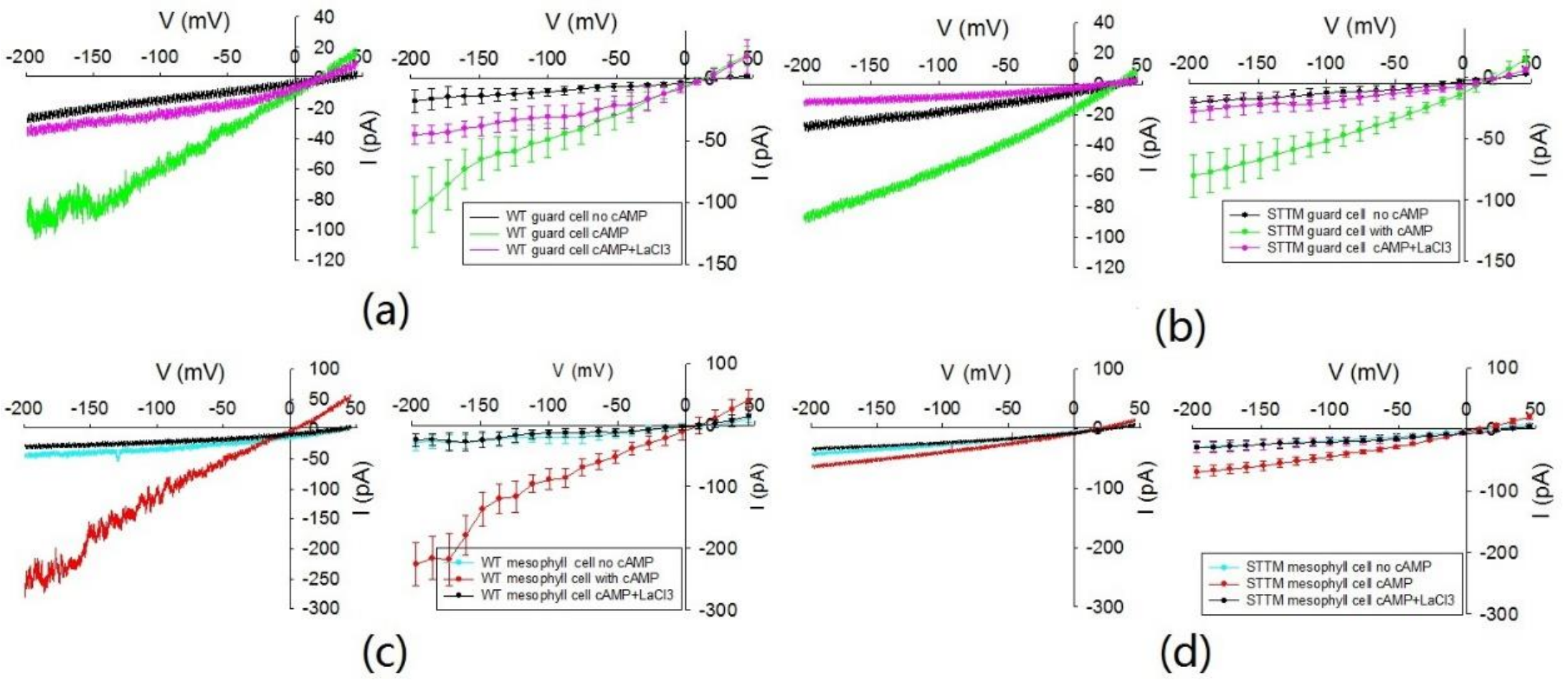

Figure 2. Reduced cAMP-Activated $Ca^{2+}$ Permeability in STTM 165/166 Guard Cell and Mesophyll Cell Protoplasts. (a) A typical recording in wild-type (WT) guard cell protoplasts are shown when there was no cAMP, cAMP, and calcium channel inhibitor $LaCl_3$ in the bath solution (left), and the statistical result is shown on the right (n=4). (b) A typical recording and statistical result in STTM 165/166 guard cell protoplasts are shown (n=5). (c) A typical recording and statistical result in WT mesophyll cell protoplasts are shown (n=5). (d) A typical recording and statistical result in STTM 165/166 mesophyll cell protoplasts are shown (n=5).

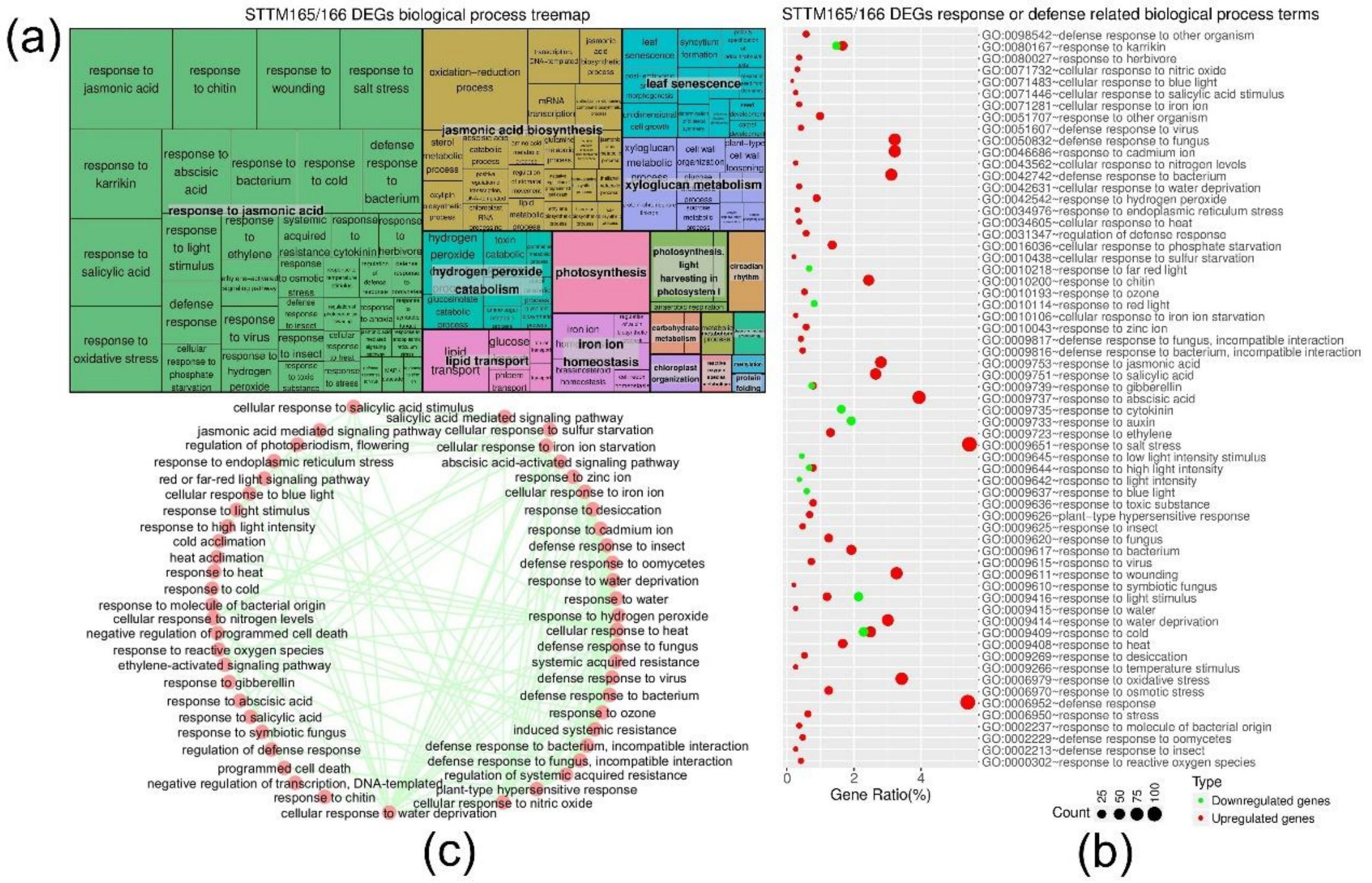

Figure 3. STTM165/166 Differentially Expressed Genes Show Enrichment of Environmental Response and Defense-Related Biological Processes in Gene Ontology Analysis. (a) A tree map showing the Gene Ontology (GO) biological processes of STTM165/166 differentially expressed genes (DEGs). (b) A comparison of response or defense-related GO biological process terms between upregulated DEGs and downregulated DEGs. (c) An interactive relation among different response or defense-related GO terms.

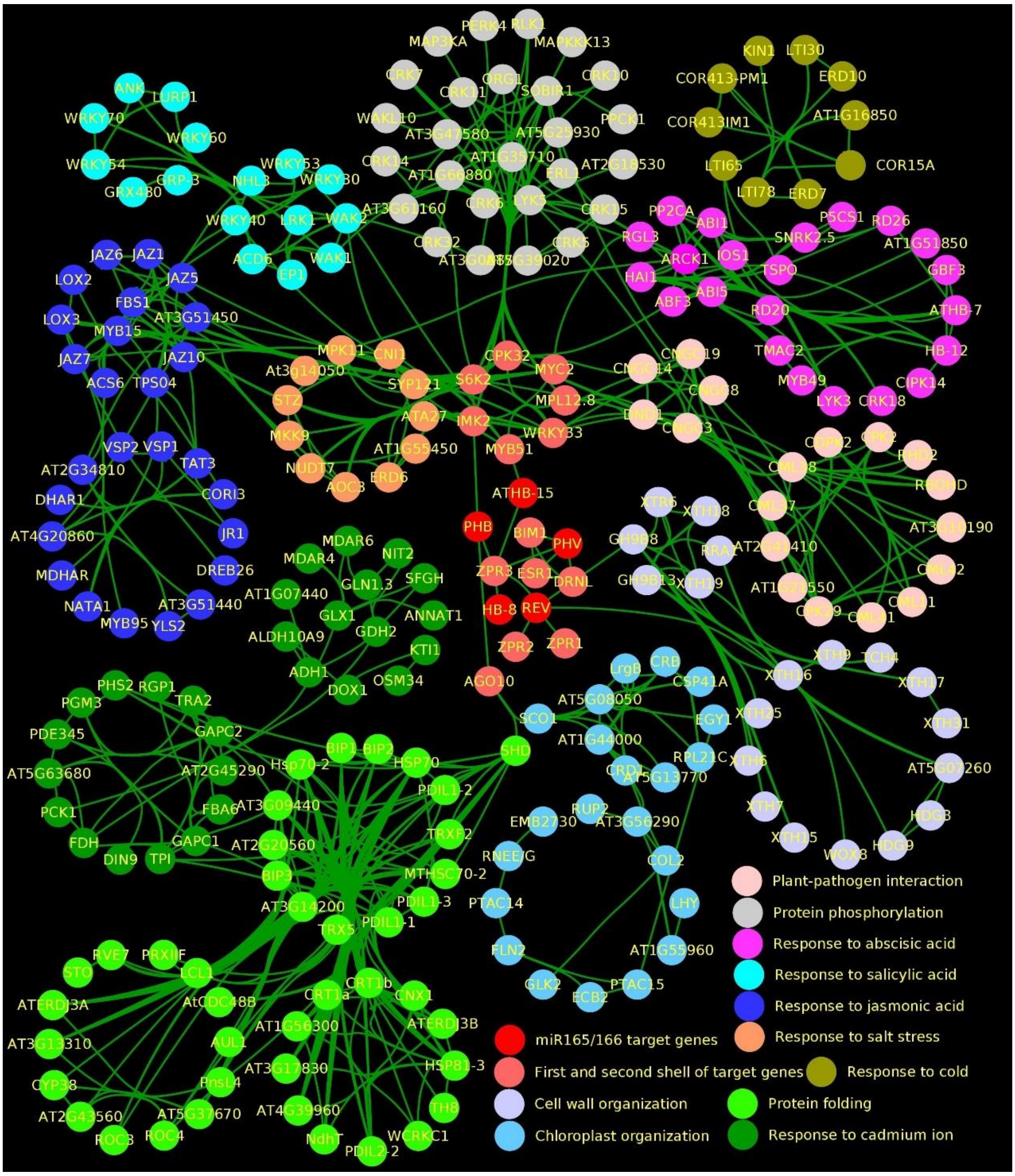

Figure 4. GO Analysis Shows Mir15/166 Target Genes in a Dominant Position within the Differentially Expressed Gene Regulatory Network. The genes directly associated with the target gene are referred to as the first and second shell genes. These shell genes are connected to several gene groups with different colors, such as those involved in the response to salt, response to jasmonic acid (JA), response to abscisic acid (ABA), response to cold, etc. These gene groups are marked in the network.

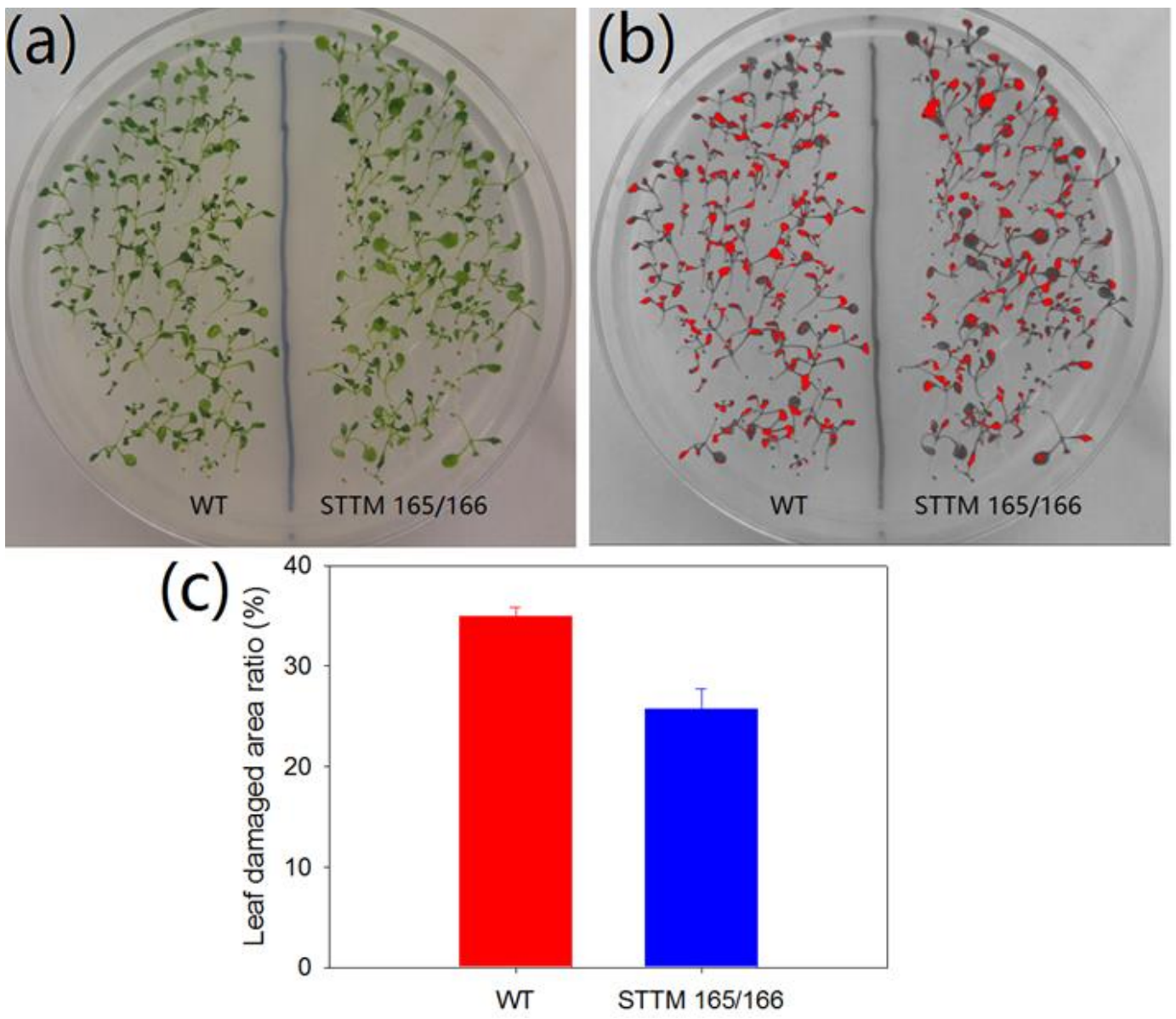

Figure S1 STTM165/166 has higher cold stress tolerance. (a) STTM165/166 and WT seeding were grown at low temperatures (4 ℃). (b) After 24 hours, the leaves' damage area was dsitingguishen and marked by red color. (3) The leaf damage area statistics.

Table S1 RNA-seq data of GLR family genes

| test_id | gene | Description | sample | sample | status | value_1 | value_2 | log2(fo | test_s | p_value | q_value | Fold Char | Diff Exp |
|---|---|---|---|---|---|---|---|---|---|---|---|---|---|
| AT4G | GLR3.2 | glutamate re | G166 | WT | OK | 5.2311 | 6.819 | 0.382 | 1.28 | 0.03 | 0.10605 | 1.303556 | NO |
| AT5G | GLR2.1 | glutamate re | G166 | WT | OK | 1.6158 | 1.8998 | 0.234 | 0.57 | 0.3584 | 0.580144 | 1.175742 | NO |
| AT3G | GLR1.4 | glutamate re | G166 | WT | OK | 2.9665 | 3.3276 | 0.166 | 0.5 | 0.4308 | 0.646624 | 1.121744 | NO |
| AT2G | GLR3.7 | glutamate re | G166 | WT | OK | 6.9057 | 7.6571 | 0.149 | 0.55 | 0.3762 | 0.596913 | 1.108798 | NO |
| AT1G | GLR3.3 | glutamate re | G166 | WT | OK | 10.938 | 11.744 | 0.103 | 0.41 | 0.5192 | 0.716982 | 1.073634 | NO |
| AT2G | GLR3.5 | glutamate re | G166 | WT | OK | 2.9319 | 2.8995 | -0.02 | -0.04 | 0.9391 | 0.97123 | 0.988928 | NO |
| AT1G | GLR3.4 | glutamate re | G166 | WT | OK | 18.468 | 18.158 | -0.02 | -0.09 | 0.8811 | 0.944061 | 0.983232 | NO |
| AT2G | GLR2.7 | glutamate re | G166 | WT | OK | 5.0239 | 4.6435 | -0.11 | -0.37 | 0.5349 | 0.729505 | 0.92427 | NO |
| AT3G | GLR3.6 | glutamate re | G166 | WT | OK | 4.3443 | 3.5054 | -0.31 | -1 | 0.1025 | 0.258407 | 0.806897 | NO |
| AT3G | GLR1.1 | glutamate re | G166 | WT | OK | 3.8855 | 2.6804 | -0.54 | -1.58 | 0.0125 | 0.054833 | 0.689835 | NO |
| AT5G | GLR1.2 | glutamate re | G166 | WT | OK | 0.2936 | 0.1671 | -0.81 | -0.9 | 0.109 | 0.268766 | 0.56921 | NO |
| AT5G | GLR1.3 | glutamate re | G166 | WT | OK | 0.7408 | 0.3899 | -0.93 | -1.52 | 0.0182 | 0.073036 | 0.526326 | NO |
| AT2G | GLR2.2 | glutamate re | G166 | WT | NOT | 0.0722 | 0.1076 | 0.576 | 0 | 1 | 1 | 1.490718 | NO |
| AT2G | GLR2.3 | glutamate re | G166 | WT | NOT | 0.0425 | 0.0457 | 0.104 | 0 | 1 | 1 | 1.07512 | NO |
| AT5G | ATGLI | glutamate re | G166 | WT | NOT | 0.0633 | 0.0376 | -0.75 | 0 | 1 | 1 | 0.594602 | NO |

Table S2 RNA-seq data of $Cl^-$ family genes

| test_id | gene | Descript | sample | samp | status | value_1 | value_2 | log2(fol | test_st | p_valu | q_value | Fold Ch | Diff Exp |
|---|---|---|---|---|---|---|---|---|---|---|---|---|---|
| AT5G24 | SLAH | SLAC1 h | G166 | WT | OK | 22.045 | 74.342 | 1.7538 | 6.3723 | 5E-05 | 0.0006 | 3.3724 | YES |
| AT1G62 | SLAH | S-type a | G166 | WT | OK | 0.0177 | 0.3181 | 4.1637 | 2.1827 | 0.096 | 0.2474 | 17.923 | NO |
| AT4G27 | SLAH | SLAC1 h | G166 | WT | OK | 2.8665 | 3.788 | 0.4022 | 1.0217 | 0.121 | 0.288 | 1.3215 | NO |
| AT5G20 | ANTI | Probable | G166 | WT | OK | 25.979 | 30.103 | 0.2126 | 0.8775 | 0.162 | 0.3501 | 1.1587 | NO |
| AT3G46 | ANTI | Probable | G166 | WT | OK | 15.057 | 15.843 | 0.0734 | 0.2511 | 0.655 | 0.8166 | 1.0522 | NO |
| AT1G12 | SLAC | Guard ce | G166 | WT | OK | 3.671 | 3.5411 | -0.052 | -0.066 | 0.911 | 0.9594 | 0.9646 | NO |
| AT2G38 | ANTI | Probable | G166 | WT | OK | 1.0918 | 1.0032 | -0.122 | -0.208 | 0.733 | 0.8633 | 0.9189 | NO |
| AT5G44 | ANTI | Probable | G166 | WT | OK | 4.7611 | 4.0395 | -0.2371 | -0.709 | 0.264 | 0.4779 | 0.8484 | NO |
| AT1G62 | SLAH | SLAC1 h | G166 | WT | OK | 0.3761 | 0.2839 | -0.4057 | -0.513 | 0.458 | 0.6684 | 0.7549 | NO |

Table S3 RNA-seq data of $H^{+}$ family genes

| test_id | gene | Descripti | sample | samp | statu | value_1 | value_2 | log2(fol | test_s | p_value | q_valu | Fold Ch | Diff Exp |
|---|---|---|---|---|---|---|---|---|---|---|---|---|---|
| AT1G6 | VHA- | V-type pr | G166 | WT | OK | 55.803 | 30.821 | -0.856 | -4.45 | 5E-05 | 6E-04 | 1.8106 | YES |
| AT1G7 | VHA- | V-type pr | G166 | WT | OK | 245.98 | 181.37 | -0.44 | -1.48 | 0.007 | 0.035 | 0.7374 | NO |
| AT4G2 | VHA-G2 | | G166 | WT | OK | 55.227 | 40.196 | -0.458 | -2.3 | 0.0056 | 0.029 | 0.7278 | NO |
| AT4G2 | VHA-G3 | | G166 | WT | OK | 0.3512 | 0.7051 | 1.0055 | 1.426 | 0.2237 | 0.431 | 2.0076 | NO |
| AT4G3 | VHA- | vacuolar- | G166 | WT | OK | 133.59 | 157.56 | 0.2381 | 1.181 | 0.1088 | 0.268 | 1.1795 | NO |
| AT2G2 | VHA- | vacuolar | G166 | WT | OK | 13.562 | 13.908 | 0.0364 | 0.148 | 0.8149 | 0.909 | 1.0255 | NO |
| AT4G3 | VHA- | vacuolar | G166 | WT | OK | 106.04 | 106.97 | 0.0126 | 0.051 | 0.9361 | 0.97 | 1.0087 | NO |
| AT2G2 | VHA-c"2 | | G166 | WT | OK | 40.421 | 39.793 | -0.023 | -0.11 | 0.8917 | 0.95 | 0.9845 | NO |
| AT3G2 | VHA-D1 | | G166 | WT | OK | 99.49 | 95.02 | -0.066 | -0.3 | 0.6511 | 0.814 | 0.9551 | NO |

Table S4 RNA-seq data of $K^{+}$ family genes

| test_id | gene | Descri | sample_ | sampl | status | value_ | value_2 | log2(fo | test_st | p_valu | q_valu | Fold Ch | Diff Exp |
|---|---|---|---|---|---|---|---|---|---|---|---|---|---|
| AT4G3 | POT13 | | G166 | WT | OK | 21.831 | 32.215 | 0.561 | 2.2864 | 0.0003 | 0.002 | 1.4757 | NO |
| AT5G5 | TPK1 | thiami | G166 | WT | OK | 22.636 | 31.211 | 0.463 | 1.7466 | 0.003 | 0.018 | 1.3789 | NO |
| AT2G3 | POT11 | | G166 | WT | OK | 24.678 | 18.138 | -0.444 | -1.674 | 0.0034 | 0.02 | 0.735 | NO |
| AT4G2 | AKT2 | | G166 | WT | OK | 15.763 | 11.521 | -0.452 | -1.842 | 0.0034 | 0.02 | 0.7309 | NO |
| AT3G0 | SKOR | potass | G166 | WT | OK | 0.4168 | 1.0498 | 1.333 | 1.6118 | 0.012 | 0.053 | 2.519 | NO |
| AT4G0 | TPK5 | | G166 | WT | OK | 4.2101 | 5.3292 | 0.34 | 1.0394 | 0.14 | 0.318 | 1.2658 | NO |
| AT2G3 | POT1 | | G166 | WT | OK | 14.349 | 16.978 | 0.243 | 0.9752 | 0.1236 | 0.292 | 1.1832 | NO |
| AT2G4 | POT2 | | G166 | WT | OK | 33.649 | 39.622 | 0.236 | 0.912 | 0.1069 | 0.265 | 1.1775 | NO |
| AT4G2 | POT3 | | G166 | WT | OK | 20.444 | 21.501 | 0.073 | 0.2773 | 0.6266 | 0.796 | 1.0517 | NO |

Table S5 RNA-seq data of $K^{+}$ family genes

| test_id | gene | Descri | sample_ | sampl | status | value_1 | value_2 | log2(fo | test_st | p_valu | q_valu | Fold Ch | Diff Exp |
|---|---|---|---|---|---|---|---|---|---|---|---|---|---|
| AT5G5 | AHA3 | | G166 | WT | OK | 51.528 | 88.7 | 0.784 | 3.0687 | 5E-05 | 6E-04 | 1.7214 | YES |
| AT5G6 | AHA11 | | G166 | WT | OK | 37.585 | 60.604 | 0.689 | 2.1843 | 2E-04 | 0.001 | 1.6125 | YES |
| AT3G4 | AHA8 | | G166 | WT | OK | 3.3097 | 5.2361 | 0.662 | 2.0987 | 0.001 | 0.008 | 1.582 | YES |
| AT2G2 | AHA5 | | G166 | WT | OK | 8.4173 | 12.392 | 0.558 | 2.1558 | 5E-04 | 0.004 | 1.4722 | NO |
| AT3G6 | AHA7 | | G166 | WT | OK | 1.123 | 1.6835 | 0.584 | 1.2553 | 0.027 | 0.097 | 1.4992 | NO |
| AT1G1 | AHA10 | autoin | G166 | WT | OK | 0.7383 | 0.7411 | 0.006 | 0.0102 | 0.987 | 0.994 | 1.0038 | NO |
| AT4G3 | AHA2 | | G166 | WT | OK | 100.5 | 85.643 | -0.231 | -0.837 | 0.133 | 0.308 | 0.8521 | NO |
| AT2G1 | AHA1 | | G166 | WT | OK | 399.57 | 333.94 | -0.259 | -0.819 | 0.174 | 0.367 | 0.8357 | NO |
| AT1G8 | AHA9 | | G166 | WT | NOTE | 0.1022 | 0.1903 | 0.897 | 0 | 1 | 1 | 1.8616 | NO |